# Assignment of a Synthetic Population for Activity-based Modelling Employing Publicly Available Data

**Serio Angelo Maria Agriesti – Corresponding author**
Aalto University, Finland & FinEst Centre for Smart Cities, Estonia
serio.agriesti@aalto.fi

**Claudio Roncoli**
Aalto University, Finland
claudio.roncoli@aalto.fi

**Bat-hen Nahmias-Biran**
Ariel University, Israel
bathennb@ariel.ac.il

**Abstract:** Agent-based modelling has acquired the spotlight in the transportation domain both in scientific literature and in real life applications. The merit goes to its capability to deal with the ever-growing complexity of transportation systems, including future disrupting mobility technologies and services such as automated driving, Mobility as a Service, and micromobility. Different software emerged, dedicated to the simulation of disaggregate travel demand framing individual choices based on the profile of each agent. Still, the actual research works exploiting these models is scarce and professionals with the knowledge to use them are few. This may be ascribed to the large amount of needed input data or to a scarcity of commercial solutions and of research production detailing the process leading to the actual simulations. In this paper, a methodology to spatially assign a synthetic population by exploiting publicly available aggregate data is proposed, providing a systematic approach for an efficient treatment of the data needed for activity-based demand generation. The methodology is implemented and validated in a case study for the city of Tallinn, Estonia. Finally, the resulting dataset, representing the whole synthetic population of Tallinn, is described so that it may be exploited by fellow researchers. Both the tools needed for spatial assignment and the resulting dataset are made available as open source.

## Introduction

In urban areas in the United States alone, in 2017, congestion resulted in 8.8 billion extra hours of travel time, purchase of extra 3.3 billion gallons of fuel, and costed overall 166 billion dollars (*1*). A similar picture appears in Europe, where congestion cost is estimated around 110 billion euros per year in terms of delay (*2*), while urban mobility accounts for 40% of all $CO_2$ emissions from road transport (*3*). The picture is also posed to become more complex in the following decades, with trends such as urbanization common to most of the globe (*4*). Indeed, an increased population entails issues such as urban sprawling, further load on the transport network, and increased levels of emissions. Different smart mobility solutions to these problems are currently being developed and tested in urban areas, with the concept of Smart City becoming more and more common (*5*). Still, new solutions often require new assessment tools and smart mobility ones are no exception. The higher flexibility enabled by digitalization requires that a certain level of disaggregation is captured by assessment models and tools, which cannot be framed by the traditional macroscopic transport models, i.e., models that consider aggregate people/vehicle flows. This prompted a surge of activity-based and Agent-Based Models (ABMs) resulting in an increase of data reliance and complexity, which is hindering their uptake and slowing down research (*6*).

Through ABMs it is possible to simulate mobility decisions down to the individual (agent) level. This allows to model and forecast travel behaviour that is sensitive to socio-economic features and individual's characteristics, to a level that is not achievable using the traditional four-step models. Moreover, the agents make travel demand choices based on the transport supply performance which makes possible, for example, to implicitly generate induced demand (*7*). These features were relevant already before the Covid-19 crisis and it is fair to assume that the pandemic will speed up changes in mobility behaviour. Indeed, in several places, historical traffic patterns have been disrupted so that remote working and higher flexibility in commuting times will require tools able to frame transport choices to the individual level. Besides, more flexible and complex mobility solutions are being developed or already deployed in cities across the globe, such as, e.g., automated vehicles, Mobility as a Service – MaaS – applications, and micromobility solutions. The effects of each one of the above, as well as of combinations of them, on transport demand is not yet univocally framed, which in turn makes aggregate models sub-optimal for assessment and future predictions.

A significant limit of ABMs is how much data reliant they are and how complex their setting is. The population of agents in an ABM includes all the residents in the study area and details some relevant characteristics of each individual, such as household structure, age, gender, employment status, etc. This kind of dataset is almost never available due to more than legit privacy concerns[1] and should be built from mobility surveys and relevant statistics (including, e.g., national census). While tools to build these datasets already exist and will be reviewed in the next Section, usually the output still does not reach the degree of precision required by most advanced ABMs and, when it does, it is thanks to the exploitation of additional methodologies and seldom available data sources. Said precision varies depending on the application; for example, in urban context, it should at least reach a feasible walking distance, e.g., a spatial disaggregation of 500x500 m. These limitations risk to stall the uptake of ABMs, which instead is to be fostered both in academia and in more professional settings.

---

[1] https://www.fsd.tuni.fi/en/services/data-management-guidelines/anonymisation-and-identifiers/

This paper aims at filling a gap in literature concerning the assignment of a synthetic population to a grid of arbitrary dimensions, namely, to a resolution of choice. This is achieved by a novel methodology for the synthetic population assignment that employs publicly available aggregate data and exploits NACE[2] margins to keep consistency in residence-workplace patterns. The result is a systematic approach, i.e., which does not leave any passage undetailed, highly replicable and designed to be flexible and nimble. In particular, the level of disaggregation can be freely designed, while the code implementing this methodology is provided as open source and customisable as needed.

The paper is structured as follows. In Section 2, we review existing literature on methods about generating and assigning synthetic populations considering, in particular, works relying on publicly available data. Then, in Section 3, the proposed methodology is described and detailed, followed by a case study implemented for the city of Tallinn, Estonia. The resulting dataset is presented and validated against the available statistical distributions in Section 4. Finally, in Section 5 we comment the obtained results and discuss future research directions.

## 1. Literature review

In this Section, we provide an overview of the state of the art concerning the generation and assignment of a synthetic population. The aim is to highlight how available methods and tools do not necessarily produce results at a desirable spatial disaggregation level while employing publicly available data. In fact, as it will be showed, the studies in literature that achieve high level of disaggregation exploit data that is rarely publicly available, such as, e.g., Origin-Destination-Industry matrixes. This, the authors argue, calls for a method that exploits public data or some aggregate statistics to carry out the spatial assignment linking synthetic populations with ABMs.

ABMs need a dataset representing the population living, working, studying, and, more important, travelling within the modelled area. Various methods and tools have been proposed to build a statistically representative synthetic version of the real population. One of such tools is the Iterative Proportional Updating (IPU) procedure introduced in (*8*). IPU processes a sample of disaggregate individuals and aggregate distributions of relevant features at certain geographical resolutions to build a synthetic population; as an example, the case study in (*8*) produced a population of 4.5 million individuals for the Great Munich area. However, as in most cases, not all variables were available as inputs and variables such as workplace or residence were allocated through Monte Carlo sampling; still, this process is not detailed in the paper. The work in (*9*) synthesizes individuals and households with a Fitness Based Synthesis algorithm instead, building a case study in Atlantic Canada. The input data were the Canadian Census and the Canadian Census Hierarchical Public Use Microdata File. The obtained synthetic population is characterized by the following variables: gender, age, ethnicity, immigrant, citizenship, household size, tenure, dwelling type, and household income. Still, these variables are too few to satisfy ABM requirements, since relevant ones, such as residence, workplaces, but also employment sector and/or status, are missing. Another tool to generate synthetic populations from travel diaries and census margins is SimPop, proposed in (*10*), which allows approaches such as model-based generation and calibration through, among others, simulated annealing. Moreover, SimPop can be utilized for very large-scale scenarios, even at country level; e.g., in (*10*) it is shown to generate synthetic population of the whole Austria (*10*). Still, no anchor point, i.e., a key place in the life and schedule of an individual, is produced when it comes to workplaces or education.

---

[2] https://ec.europa.eu/competition/mergers/cases/index/nace_all.html

Information about spatial patterns of housing is considered in (*11*), where a two level Iterative Proportional Fitting (IPF) method is applied, which assigns residents to each building while exploiting data such as dwelling type and household income. As detailed above, this degree of precision is actually rare in literature; however, this comes at the cost of utilizing data that is rarely available, such as the average transaction prices or building capacities. Moreover, in this case, the degree of precision may be unnecessarily high for ABM setting in the context of transport demand modelling. Finally, existing literature suggests that IPU outperforms IPF in term of generation of a population, since the latter does not allow to match the joint distribution both at the individual and at the household level at the same time (*10*). Besides, further improvements of the IPU algorithm have been proposed in time. In (*12*), for example, the IPU algorithm is updated to be able to control for constraints at multiple geographic resolutions when generating a synthetic population.

In (*13*), two synthetic generation methods (one sample-based and one non sample-based) are tested on a portion of the French population; however, the focus of the paper is the comparison between the methodologies and few details about the input and the achievable output data are provided, while no ex-post spatial assignment is performed. The work in (*14*) adds land use variables in the generation of synthetic population and the results suggest that the addition of such variables improves the capability to frame additional nuances, such as, e.g., the differences in mobility patterns between rural and urban areas. In (*15*), the assignment of schools and workplaces is performed based on different assumptions to tackle the lack of data related to enrolling and commuting patterns. While for education, age, and distances were the main factor defining enrollment, the workplaces were assigned randomly within counties (still in a way to meet total margins from the census).

A different approach for the population assignment is suggested in (*16*), where a synthetic population of establishments (i.e., places of business) is built first for an ABM model (the SimMobility platform). The authors then suggest that such results may be exploited to assign anchor points to the population, since all the relevant information are modelled. Still, some of the required data are rarely available, such as, e.g., the establishment locations, industry type, employment size, and occupied floor area. In (*17*), different prototype cities, populations, and mobility patterns are built to create different scenarios; in particular, the prototype population is constructed by assigning spatial features based on land use characteristics. However, the paper neither explores patterns between residence and workplace location, nor details how a (gravitational) model is applied to fill this gap. It is then left unclear how the solution actually reproduces commuting patterns, as well as if an excessive skeweness in indicators other than the total margins could be introduced through it; moreover, since the assignment of the synthetic population is not at the core of *(17)*, their methodology is not detailed.

Another recent study tackling the workplace assignment issue for synthetic populations is (*18*), in which an origin-destination-industry matrix is exploited to assign workplace probabilities to the synthetic population for the Greater Boston Area. While this approach allows exploiting observed patterns rather than assuming theoretical models, such origin-destination-industry matrixes are rarely available. Similarly, (*19*) exploits data such as commuting patterns, commuting OD matrices and distance travelled to carry out the assignment of workplaces for a synthetic population through multinomial distribution.

A good overview of the current state of the art is given in (*20*) which expands the state of the art by building both a synthetic population and the arising travel demand with only open and publicly available data. This also makes a good case about why research on this area is needed, namely how other current approaches are rarely systematically tested and hardly transferable. While the aim and contribution are similar, some challenges are addressed in the presented paper that are not addressed in said work. For example, in (*20*), data concerning the commuting matrixes were openly available and could be used as proxies. This is not the case in other locations (including Estonia), where other statistical information (e.g., the NACE margins) should be exploited instead. Besides, information related to businesses has a high level of detail in (20), while in this paper again statistical information and land use data are exploited as proxies to assign workplaces. Finally, the exploited household travel survey had enough entries to allow correlations between socioeconomic features and commuting distances/anchors (consequently, it facilitates to build the daily activity schedules for the synthetic population). This was not the case in our work. Therefore, the method reported in this paper differs from (*20*), even though both the motivation of the work and the aims are similar. Both works try to provide a runnable pipeline simple and as reproducible as possible, built only on publicly available data. To the knowledge of the authors, aside (*20*) and the presented paper, no other work investigates similar methods or challenges the state of the art in a similar way.

To summarise, in literature, the small but crucial step of assigning a synthetic population to disaggregated spatial units seems either to rely on very detailed data, e.g., about firms and commuting patterns, or to be limited to aggregate margins and probability distributions. One of the many ABM examples of the latter is (*21*), a work exploiting MatSim. The paper reports how residence and workplace locations were randomly assigned in areas with coherent land use (e.g., no residence is assigned to land use zones without residential buildings). (*21*) is very relevant as an example because it explicitly aims to use only publicly available data to build the model. Still, this solution is not feasible in cases where only very aggregate data are available (i.e., in the case study presented below, where workers totals were available only at district level).

Finally, even though some tools have embedded functions or plugins to assign anchor points (e.g., ct-ramp[3]), others do not (e.g., SimMobility MT[4]) and the latters need agile methods to implement this assignment. The proposed methodology is independent from any ABM software and can be used regardless of the model that is being exploited. The methodology presented in this paper is conceived to be both less reliant on firms' data/ commuting OD matrices and nimble enough to integrate additional inputs. Besides, it is the only solution, to the authors' knowledge, that makes up for the lack of disaggregated statistics with the introduction of NACE fields. Providing a systematic approach to tackle the challenge of assigning anchor points to a synthetic population, employing only public data, should also allow the state of the art to further move forward ABM. By exploring the literature, it is evident that *no approach allows researchers to fit synthetic populations for ABM in a formalized, efficient, and quick way while exploiting only publicly available data, aggregate margins, and land-use features, regardless of the exploited tools.*

## 2. Methodology

As previously argued, the generation of a synthetic population does not frame all the anchor points needed to design an Activity Based model. Indeed, an intermediate step is needed: the spatial assignment of the population to a level of disaggregation that is fit for the required implementation. Input of this step are the generated synthetic population and, for the proposed methodology, statistical data, such as cadastral data, industry statistics, and total employees' or students' margins, as reported in Figure 1. The output of this step is the synthetic population integrated with anchor points, input for activity-based demand generation. The outputs have been widely tested on SimMobility Preday but should fit any activity-based demand generation model.

The novel contribution of this paper focuses essentially on the highlighted part in Figure 1, namely the spatial assignment of anchor points, while assuming other state-of-the-art methods are employed for the other components.

---

[3] https://www.ct-ramp.com/

[4] https://github.com/smart-fm/simmobility-prod/wiki/Introduction-to-SimMobility

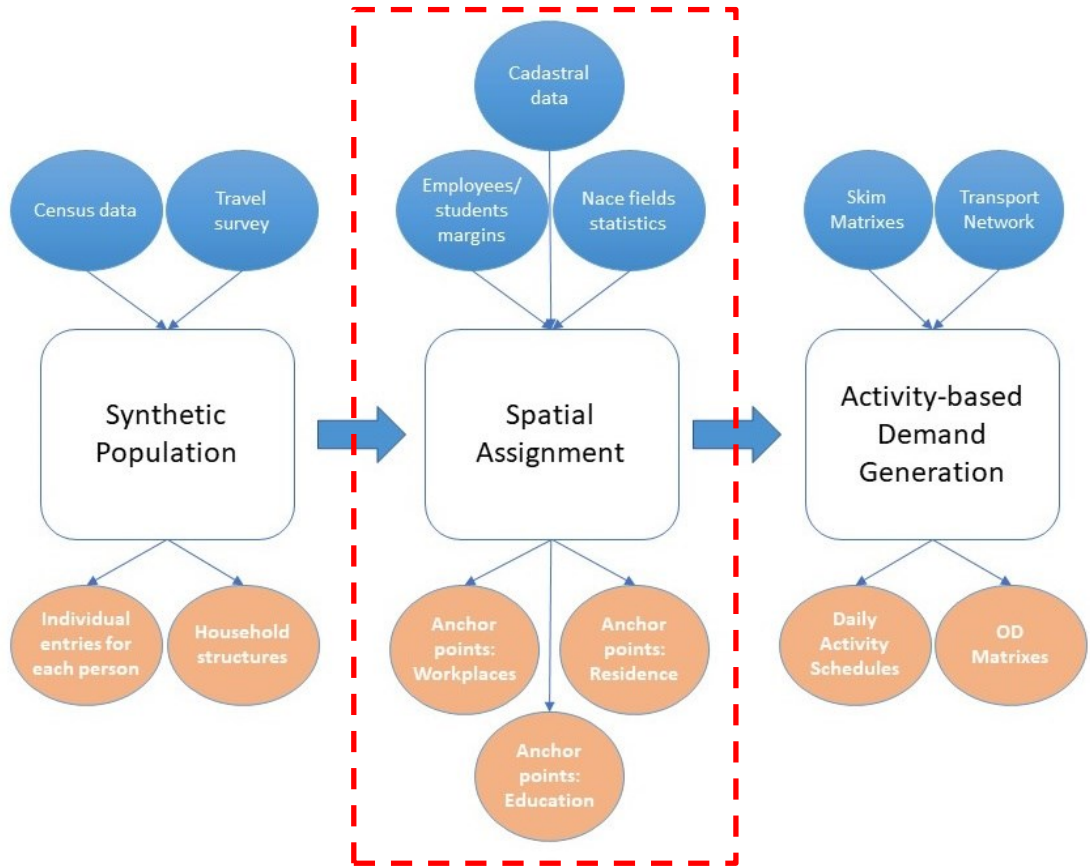

**Figure 1:** Conceptual framework: From generating the synthetic population to activity-based modeling

Before presenting the proposed methodology, we briefly summarize the most relevant method existing in literature *(17)*, which was developed for building different prototype cities with their prototype population, where spatial features are assigned based on land use characteristics and distance. The algorithm proposed in *(17)* can be summarized as follows.

First, weights for different cell classes are defined as:

$$G^{\text{work}}(g_L, g_H, g_C, g_I, g_E, g_O) = (1, 2, 10, 5, 3, 1),\tag{1}$$

where $g_X$ is the weight of class $X = \{L, H, C, I, E, O\}$, defined as: low residential ($L$), highly residential ($H$), commercial ($C$), industrial ($I$), education ($E$), and open land ($O$). Then, the cell weights are normalized within each subzone (i.e., second administrative district level), according to:

$$p_{i,s}^{\text{work}} = \frac{g_i^{\text{work}}}{\sum_{i \in C_s} g_i^{\text{work}}},\tag{2}$$

where $i$ indexes cells, $s$ indexes subzones, and $C_s$ is the set of cells within subzone $s$.

The number of workplaces in each cell within a subzone, $N_{i,s}^{\text{work}}$, is then computed by multiplying the normalized weight by the total number of workers within a subzone $N_s^{\text{work}}$:

$$N_{i,s}^{\text{work}} = p_{i,s}^{\text{work}} * N_s^{\text{work}} . \tag{3}$$

Finally, distance is used for a last mile assignment of the workplaces to the population.

This method allows to perform the synthetic population assignment relying on data that is commonly publicly available such as land use data. The degree of disaggregation that is reached is indeed sufficient to build accurate activity-based models, since the dimension of the cells can be arbitrarily defined. However, this method does not explore how a final gravitational model-based assignment is carried out for the workplace, nor if any kind of further spatial consistency or commuter patterns across subzones are considered in the workflow. Therefore, it is not clear how closely the residence-workplace relationship is matched through the presented method. By including the NACE assignment, our work tries to build on such framework to include factors other than land use and distance. Moreover, describing the gravitational assignment and the results should foster transferability and replicability of the proposed methodology.

## 2.1.   Assignment of NACE fields to the population

The first proposed step consists in exploiting census data to obtain the distribution of workers by their occupation status and their NACE classes (or their NAICS[5] ones, or similar), based on gender, age, and district of residence (as recorded from the national census[6]). The assignment is implemented through probability distributions such as *occupation per age, gender, and district of residence*. At the end of the assignment, unfeasible or very unlikely combinations, such as, e.g., 20 years old managers, should be removed, by reassigning the occupation status. This process is illustrated in Figure 2.

Once the assignment of the NACE (or equivalent) field is performed, the totals within the synthetic population is checked against the total number of employees within the city, as recorded for example by the business census. Indeed, we should consider that the synthetic population, while representative of the overall population, may misrepresent some distributions depending on the distributions it is calibrated against (*10*). This may lead to skewed totals in some NACE fields or in some districts/subzones. Moreover, it may happen that the NACE dataset and the business register are not consistent with each other.

---

[5] [North American Industry Classification System (NAICS) U.S. Census Bureau](#)

[6] The EMTAK classification, national version of the NACE classification, was exploited

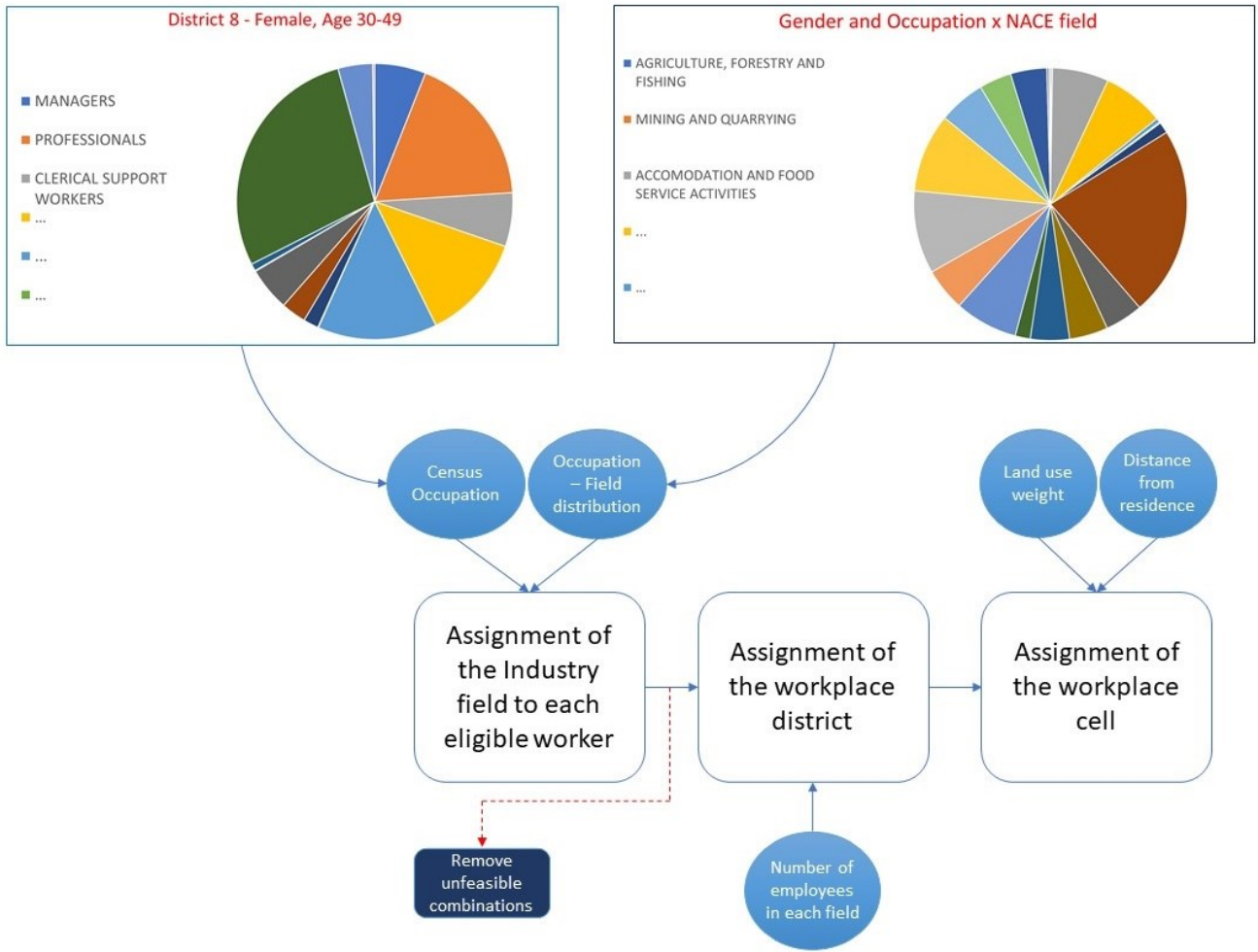

**Figure 2:** Workplace assignment - phases and example of data exploited

## 2.2.    Subzone assignment

In case no inconsistencies between datasets are detected, the district containing the workplace for each individual is assigned so that the totals per NACE field are met. The assignment is carried out extracting random samples based on the probability distribution *NACE field per district of work*. The NACE field are assigned based on age, gender, and residence through the occupation distribution. A stronger tie between these variables and the working district is therefore achieved compared to the one that would have been obtained by simply exploiting aggregates to derive a distribution while matching the total employees' margins.

When the NACE field cannot be exploited because it is inconsistent among datasets, the remaining population is clustered into the NACE field "Other". In this case, we first compute the distance between each residence cell and each other subzone ($d_{i-s}$); then the first subzone gravitational pull, namely the number of employees not already assigned to another NACE field, is considered and the probability for each individual to work in one of the districts $p_j$ is computed as:

$$p_j = \frac{w_s}{d_{i-s}},\tag{4}$$

where $j$ is the district of work, $w^s$ is the number of employees in said district, and $i$ is the cell of residence.

Despite this method may add some noise to the total number of jobs in a district, spatial integrity is preserved, i.e., people from one side of the city are less likely to work on the other side of the city. However, this step is needed to keep this assignment comparable to the one for coherent NACE fields. Moreover, as it will be showed in the case study, the number of employees assigned this way in each subzone remains fairly consistent with the real-world data.

## 2.3. Last mile assignment

Once the district is assigned, we allocate to each individual the class of the cell in which the workplace is located. This step basically guarantees that the land use distribution is not skewed by the gravitational model-based assignment that will be carried out as last step. The probability for each individual whose workplace is in the considered district to work in one of the cell classes is equal to:

$$p_{ij} = \frac{\sum g_{ij}}{\sum_x^X g_{xj}},\tag{5}$$

where $g_{ij}$ is the weight (calculated based on the prevalent land use destination within each cell, as recorded in cadastral data), $i$ is the class (highly residential, low residential, businesses and services type, and manufacturing type), $j$ is the considered district, and $X$ is the number of cells in the district. Once the cell class and the work-subzone are assigned to an individual, the work-cell is assigned purely based on the distance from the residence cell $d_{nm}$:

$$p_{m,j}^{\text{cell}} = \frac{1}{d_{nm}},\tag{6}$$

where $n$ is the cell of residence and $m$ is one of the cells *of the defined class* in district $j$.

It is worth highlighting that, until the last step, the distance factors have been used only for minor adjustments, namely a) as a proxy for unrealistic NACE distributions (calculating the average of distances between the residence cell and each cell in the target district) and b) as a corrective factor for the subzone assignment (considereing the distance between the residence cell and the cells in the target class). Still, the above passages exploit NACE fields and land use data to increase the representativeness of the assignment whenever possible, before resorting to the distance for the last mile assignment. Indeed, the field of work is bound to be a more representative variable than distance, since rarely a person has the freedom to choose her/his workplace to a 500x500 m precision, whereas factors such as salary, kind of job, etc. are much more significant. It is important to highlight that distance, especially when NACE fields are exploited, accounts only for the last mile cell assignment (in the 500x500 m grid) and only after the weight of the cells have been considered. Thus, the proposed method, by exploiting land use and NACE statistics, *does not* result in an unrealistic distribution of workers deciding to work near the place of residence. An example will be provided in Section 4.

Finally, it is worth mentioning that the proposed framework is flexible enough to allow other factors other than the employment sector to be integrated, in the case a proper data source is available, while still keeping the overall process simple and parsimonious in term of data requirements. For instance, one may include major public transport nodes in the weight computation or consider additional land use classes (e.g., an interesting work in this direction is (*22*)).

## 3.   Spatial assignment: The case study of Tallinn

### 3.1.   General description and data availability

The methodology is applied here on a case study related to the city of Tallinn, Estonia. Tallinn is a European capital characterized by an efficient bus and trolleybus network and an important port that serves high flows of both freight and commuters. It is also located on one of the TEN-T axes (Rail-Baltica).

The chosen reference year is 2015, motivated by the fact that a travel survey (*23*), which included travel diaries, was recorded during that year. In 2015, the city counted 414000 inhabitants, the most populated among the 8 districts being Lasnamäe and Mustamäe; relevant statistics and distributions such as Gender per Age were obtained both from the national statistical database (www.stat.ee) or from the municipal records (*24*).

We report in the Appendix an overview of all the available data that could be utilised. Essentially, all data sources were openly available for free, with the only exception of workplace aggregates that required a fee for the extraction process. It is worth highlighting that, as argued in Section 2, non-public data sources (e.g., mobile phone data, company addresses, floor space, disaggregate statistics, commuting ODs, etc.) or any data allowing for more complex assignments were not considered in this study.

Maps showing districts and subdistricts are provided in Figure 3.

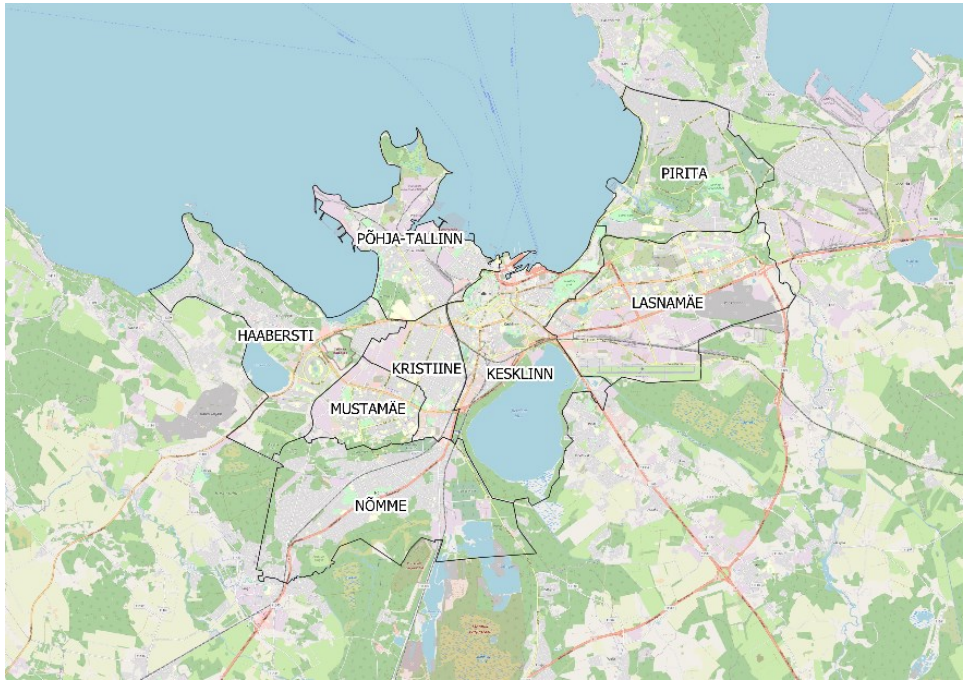

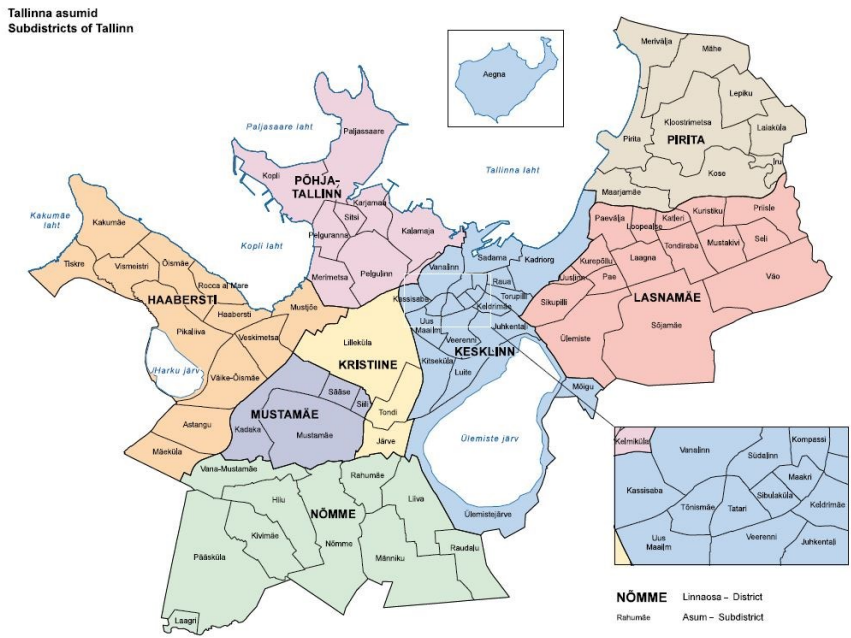

**Figure 3:** Tallinn districts (top) and subdistricts (bottom). (Source: Estonian Ministry of the interior, Population Register)

The process that was followed to generate and validate the final dataset for the city of Tallinn is summarized in Figure 4, where each step is tied to the needed inputs.

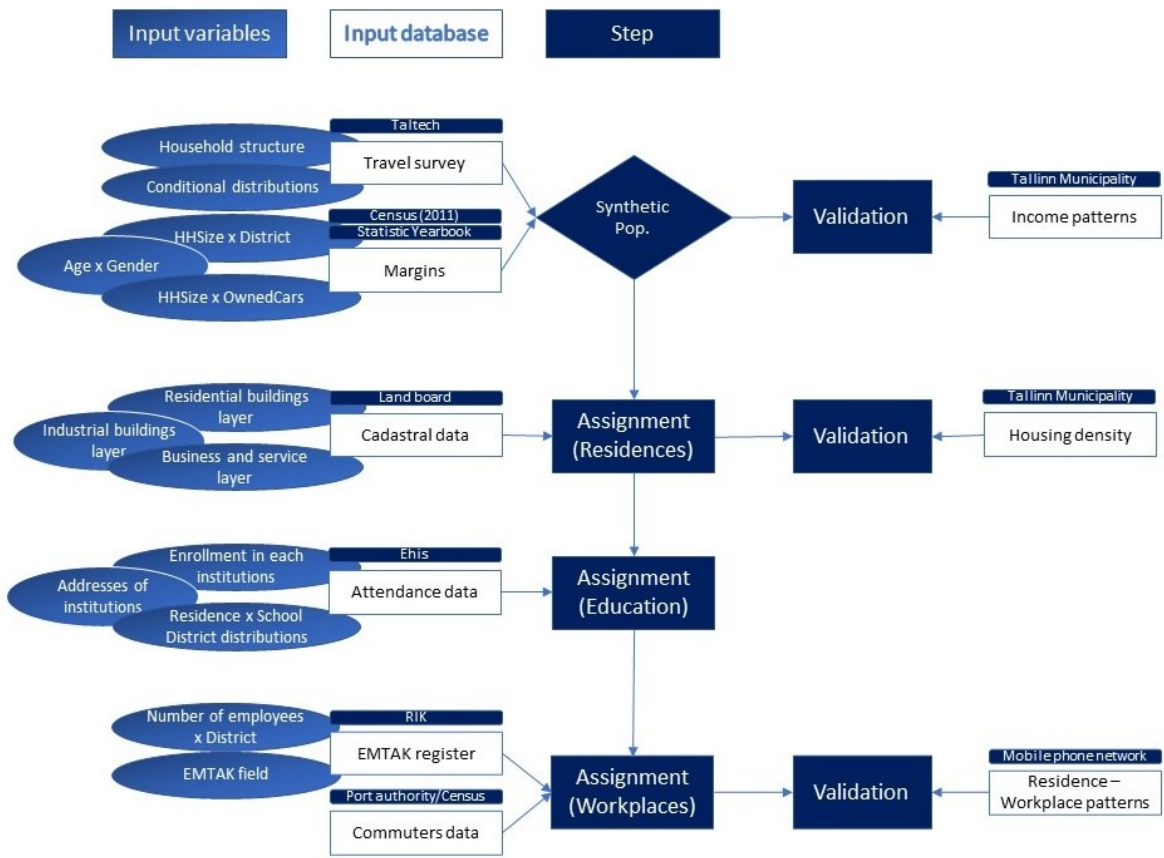

**Figure 4:** Process - From synthetic population to input dataset for an ABM

As it can be seen in Figure 4, the work on the synthetic population follows four main phases, for each of which both the sources and the exploited variables were reported. Besides, on the right, the data used for validation was listed, again highlighting the sources. Figure 4 also tries to highlight the modular nature of the process.

### 3.2. Generation of the synthetic population

In order to generate the synthetic population to be later assigned, the SimPop package was chosen (*10*). However, it must be highlighted how the choice of SimPop does not limit the applicability of the presented approach, since basically any other method able to produce a synthetic population can be exploited to generate the needed inputs. For additional details concerning the SimPop tool, interested readers may refer to (*10*). Despite this first step produces an initial version of the synthetic population, this is not yet ready to be exploited as input for an Activity-based demand model, as it still lacks key variables such as places of education and workplaces. As it can be seen from Figure 1, these variables are modelled through the intermediate step "Spatial Assignment" and result in the key anchor points needed to replicate typical travel patterns.  However, the anchor points could not be included because they were not properly captured by the travel diaries. Furthermore, the residence distribution had to be further disaggregated since each agent should be assigned to an area that would allow to model her/his mobility choices (e.g., 500x500 m). Indeed, it is not possible to apply choice models to areas as big as subdistricts while it is possible to do so when the degree of precision is within walking range.

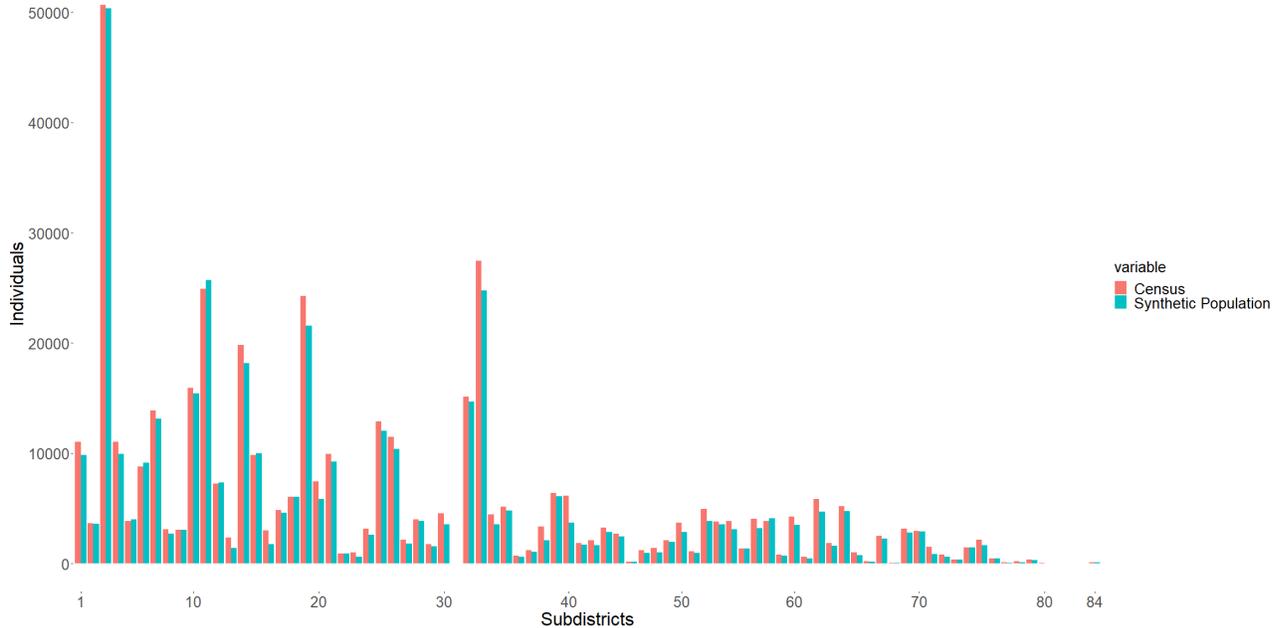

**Figure 5:** Subdistricts population resulting from SimPop

We report in Figure 5 the first version of the synthetic population resulting from applying the SimPop tool, where it may be observed that the two populations (real, i.e, from census, and synthetic) match almost perfectly in terms of household sizes and spatial distribution across Tallinn's subdistricts. This happens indeed because the final calibration was carried out through simulated annealing on these two variables (which were deemed more important for the case study than, for example, the *gender x age* distribution). In addition, due to the survey structure, the average income per family member was more closely related to the household structure than to the individual. Thus, since income is another key variable for an ABM, a more precise household size distribution was favoured over calibration on individual variables.

As it can be seen from Figure 6, the income distribution for the synthetic population matches reasonably well, at least in qualitative terms, the real one. The only inaccuracies appear for the small subdistricts, due to the very low number of residents. To maintain the synthetic dataset anonymized, the income per family member variable was converted into four levels (high, average, low and not available). The total number of cars was instead assigned based on the margins found in (*25*), following a probability distribution based on household variables.

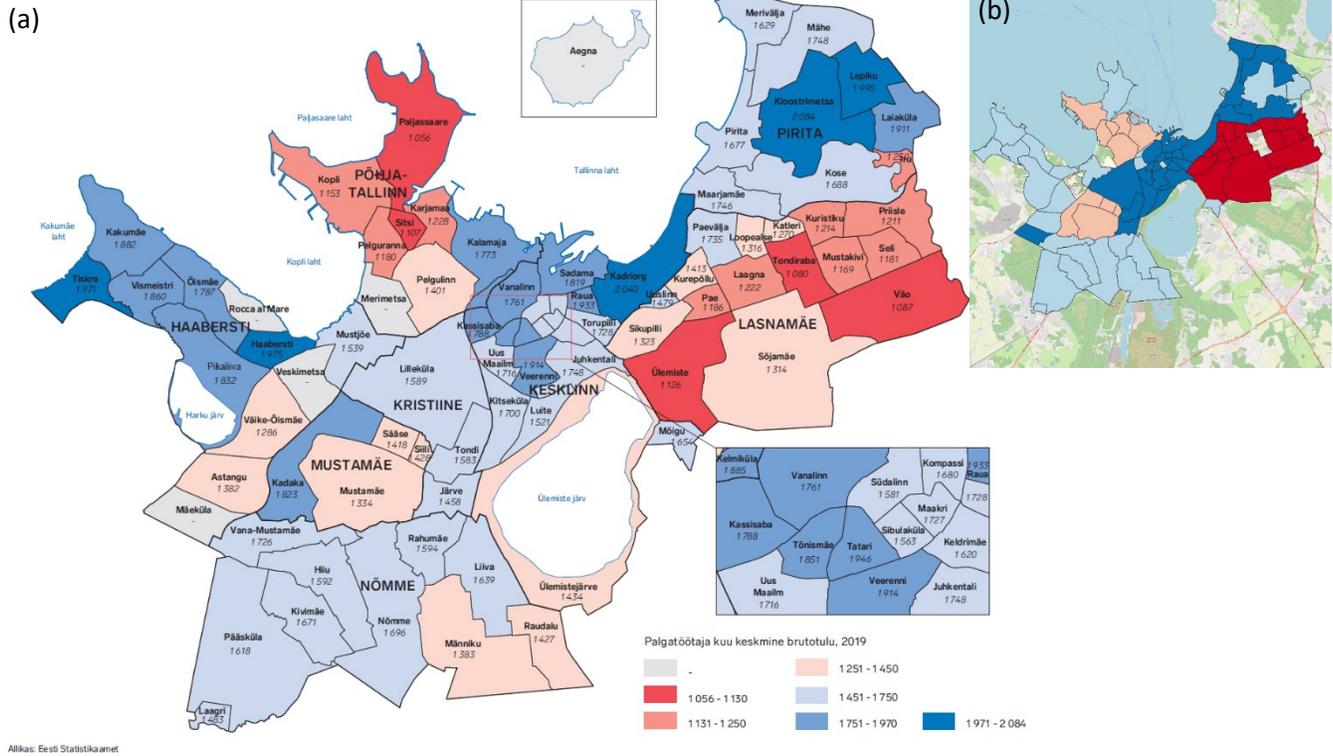

**Figure 6:** Income distribution across subdistricts:(a) Real population (Source: Eesti Statistika amet); (b) synthetic population (dark red = lowest income, dark blue = highest income)

As mentioned, activity-based models need a level of spatial disaggregation not framed by the census data and not reproducible through SimPop. Therefore, the methodology defined in (*17*) was applied. In our case study, the city of Tallinn is split in 628 500x500 m cells, which are then categorized as highly residential (HR), low residential (LR), businesses and services type (OW), and manufacturing type (MW) based on cadastral data and the different destinations for the buildings in each cell, as shown in Figure 7.

The single cell values for residents seem to precisely match the ones recorded (but not publicly shared, to the best of our knowledge) by the city census, as in Figure 8. As it may be observed, the darker blue cells in the left picture overlap with the ones in the right one.

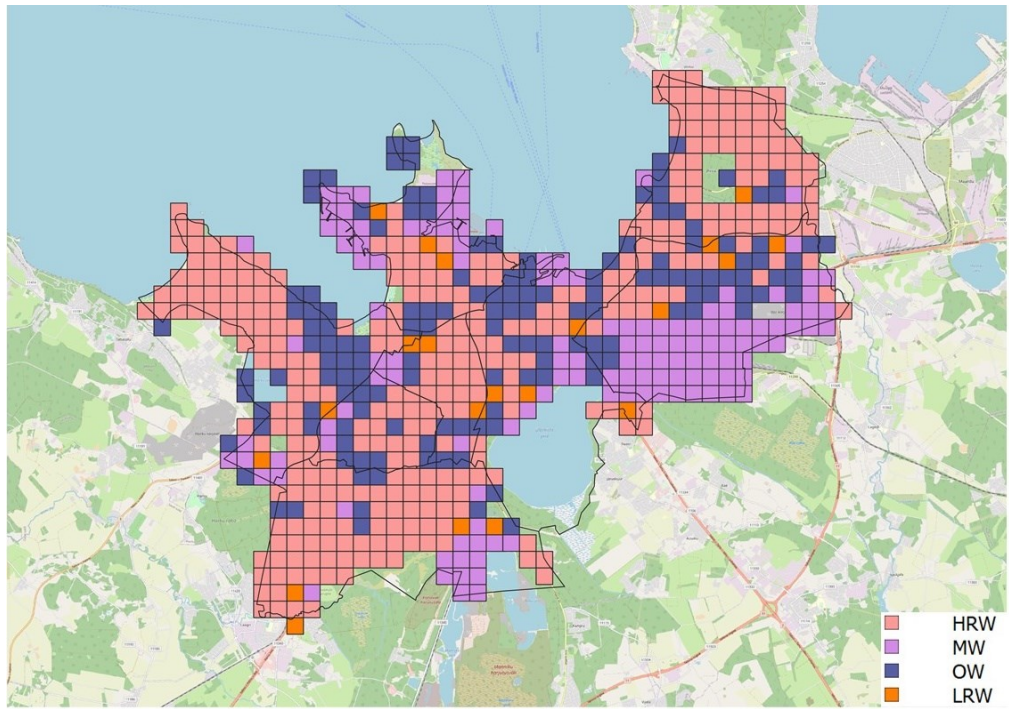

**Figure 7:** Scored grid and cell classes in the city of Tallinn

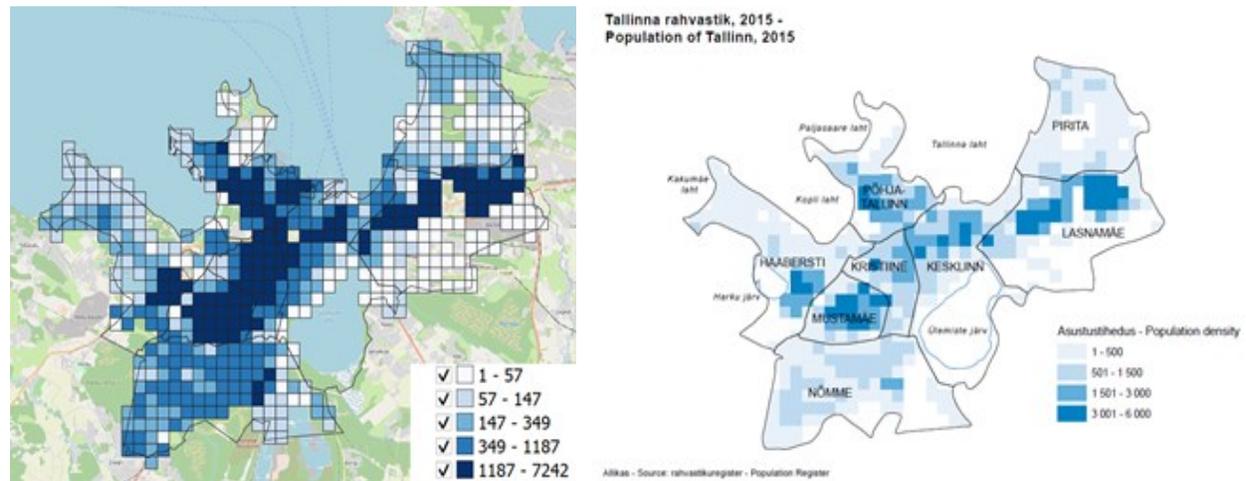

**Figure 8:** Resident distribution for the synthetic population (left) and from the census (right - source: rahvastikuregister)

## 3.3.   Spatial assignment - Workplaces

To implement the spatial assignment of workplaces, an aggregated dataset was obtained from the Estonian Centre of Registers and information Systems (RIK). Such dataset differs from the more complex ones reported in literature by being aggregated and anonymized; in fact, only the total margins for the number of employees in each EMTAK field at district level were exploited. In this section, each individual is assigned a workplace within a cell in the grid, without assigning her/him to a specific building or address. The exploited distributions are reported in Figure 9 and 10.

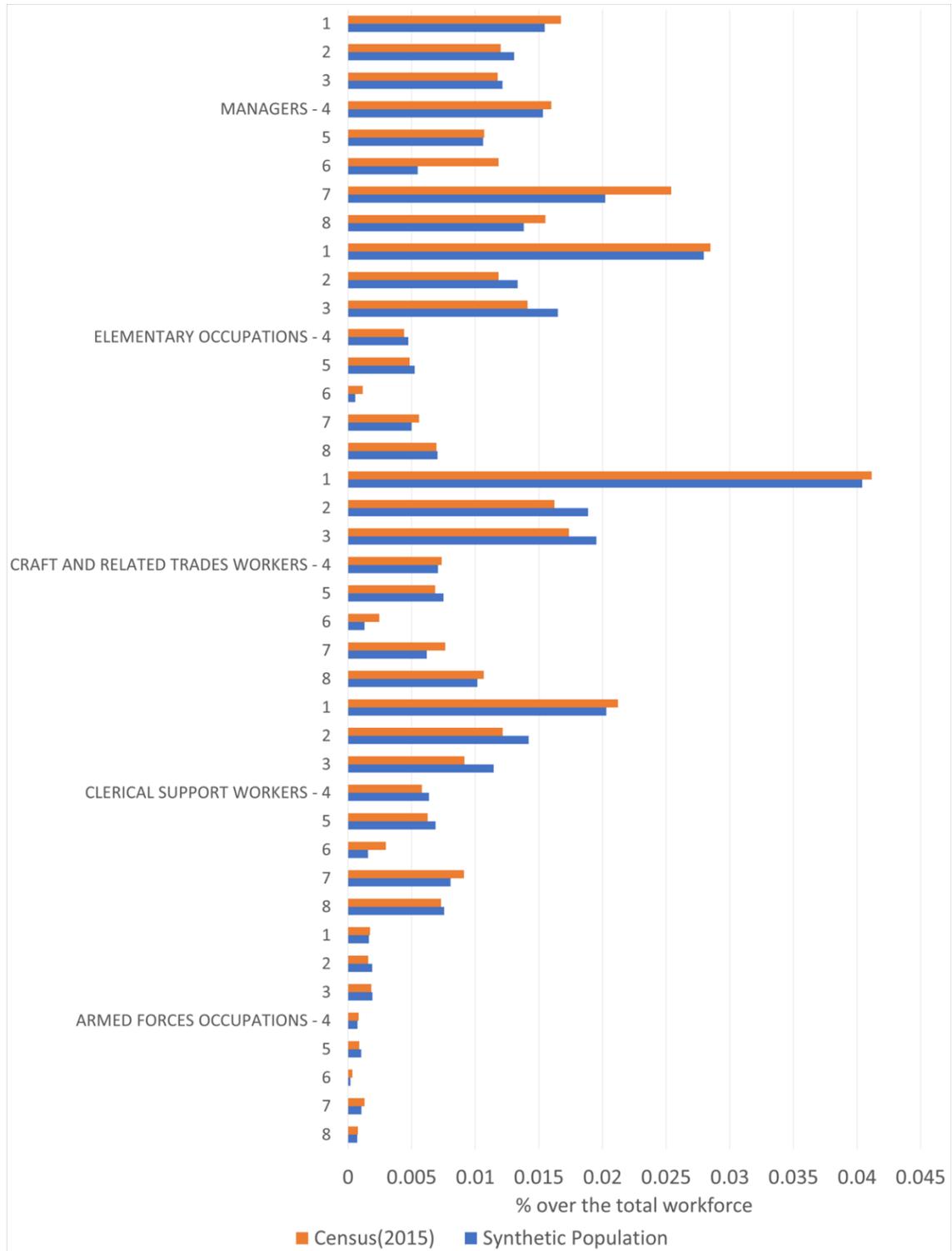

**Figure 9:** Example of occupation and spatial distribution per district of residence (numbered in the range 1 - 8)

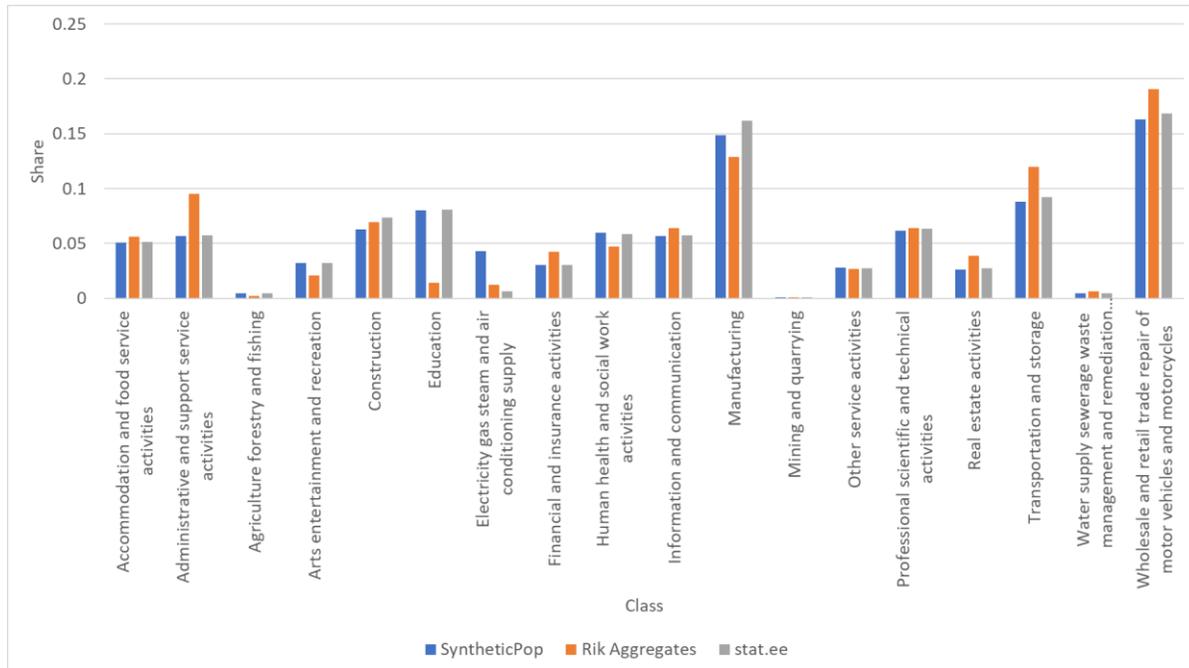

**Figure 10:** EMTAK/NACE classes: synthetic population against RIK dataset and census (source: stat.ee)

In the following, the application to the Tallinn case study using the methodology described in Section 3 is presented. First, from publicly available census data, one may obtain the distributions of workers in the EMTAK field based on occupation (which, in turn, is related to gender, age, and district of residence). This is exploited to assign a field to each individual in the synthetic population, while keeping consistency in the spatial distribution of professional roles; Figure 9 reports the achieved matches among occupation and the overall population. It must be highlighted again how the synthetic population may slightly differ from the actual one in some areas or in some features. This is due to the stochastic nature of the assignment and depending on the margins the population is calibrate against. It is therefore important to verify that the margins are consistent with the census one; in this case, the occupation fields were assigned based on age, gender, and district of residence and we can conclude that the degree of consistency is satisfactory. Once the assignment of the EMTAK fields is performed, the totals in the synthetic population arising from the census data are compared with the totals in the RIK dataset and inconsistencies are identified. For example, the workers in "Education" are 14118 in the census assignment, while they are only 2825 in the RIK dataset (see Figure 10). All the outliers are then categorized as "others" and their field is not exploited in assigning the workplace district; whereas, for the fields whose distribution matches between the two datasets, it is instead possible to exploit the distribution.

Once the EMTAK dataset is coherent among the RIK and the census databases, both the assignment of the EMTAK field and of the workplace district are carried out by applying the recorded distributions (as detailed in Section 2). Thus, the focus will be on the last mile, which is implemented as follows:

- First, weights based on the class of each cell are assigned, which allow also to calculate the total weight for all the cells in each district;
- then, the ratio $i$ between each cell weight and total of the weights in the district is calculated according to Equation 2.

For the EMTAK workers whose total are coherent between RIK data and the census, the ratio $i$ is exploited to assign the cell of the workplace. Distance and EMTAK fields are then exploited to assign the subzone, namely the work-district, which allows to frame coherent and realistic commuting patterns. For the other workers, the simple cell assignment was deemed potentially too skewed and the work-district assignment was not possible a-priori based on census or EMTAK distributions.

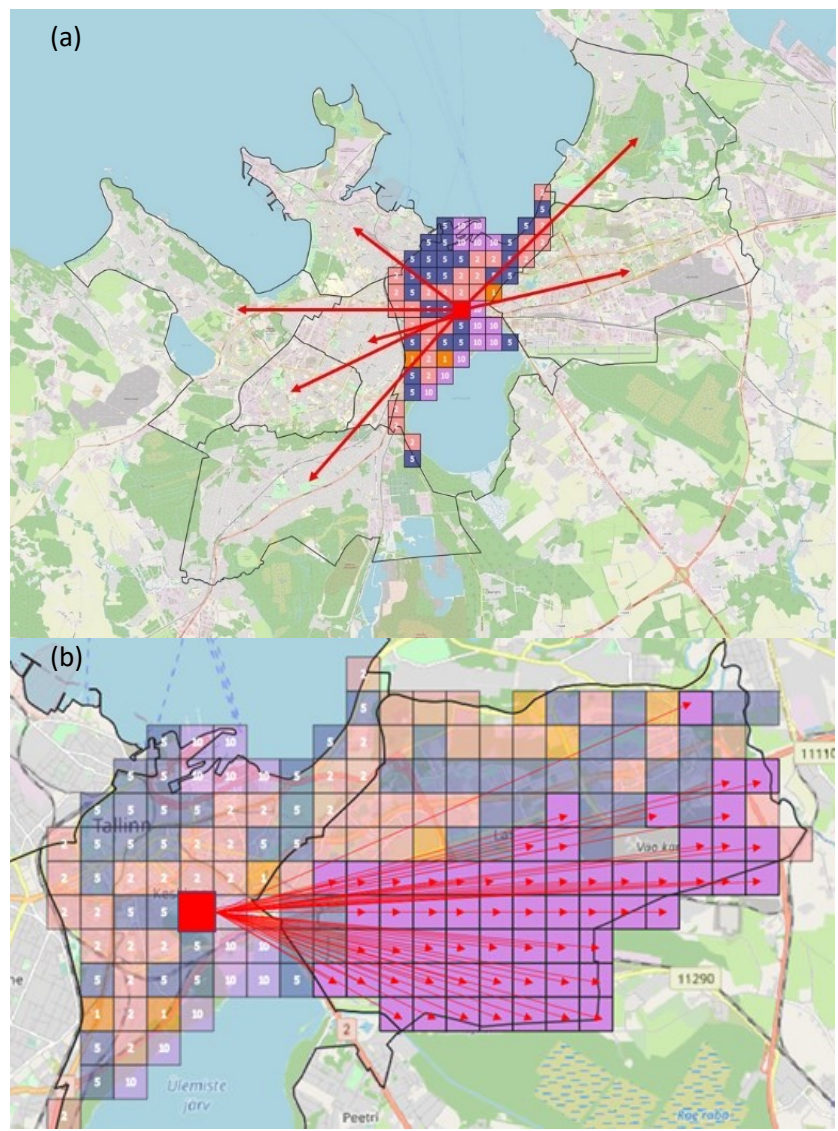

**Figure 11:** (a) Land use weights for Kesklinna and distances from each district for each cell; (b) Last mile assignment based on distance (linear)

Therefore, to avoid having individuals working on the other side of the city only based on the cell weights, the following heuristic algorithm is applied:

1. The distance between each residence cell and each district, calculated as average between all the distances between the cell at hand and the ones included in the district, is computed (see an example in Figure 11a).
2. For each cell pair (one being the residence, the other being the eligible workplace), the ratio between their distance and the average distance among the residence cell and all the other cells in the district is calculated.
3. Each district has its own gravitational pull calculated based on the number of employees in the remaining fields ("others"). In this case, the probability of working in a district is calculated via Equation 4. Even if a certain noise is added to the total number of jobs in the "other" field, a spatial integrity is kept (distant districts have less chances of being chosen). Besides, it will be showed how the total number of "other" employees in each district remains quite consistent.
4. The class of the workplace cell is calculated based on the cell classes distribution within the district via Equation 5.
5. Once both the district and the class are assigned for the workplace, the final cell assignment is simply carried out through Equation 6; Figure 11b illustrates how the final cell assignment is carried out within each class.

Note that distance as a factor is only consider as a corrective item when assigning the work-place district (the main factor being the gravitational pull of the number of jobs or the actual EMTAK distribution) and for the last mile assignment. Moreover, when assigning the work district, the distance factor is considered only for the fields of work for which no reliable spatial information was available. The above was carried out through only spatial data concerning the number of buildings in each cell and their main destination and the aggregate statistics at district level about the number of employees per EMTAK field (where possible).

Figure 11b shows the last step of the method, which involves the distance assignment. People residing in the red cell have already been assigned to the district of work based on census margins and NACE fields and to a cell class based on the relative weights and the resulting probability. Then, after the workplace has been defined as one among the purple ones, the distance is employed as final proxy. Indeed, since the purple cells fall in the manufacturing type and have a high weight, the number of workers in the distant cells on the eastern border turns out to be higher than the workers in the west and northern section of the district (the ones nearest to the red cell of residence).

### 3.4.   Validation of the resulting dataset

In the following, we present an overview of the relevant distributions and we investigate how they match real world data. As it can be seen in Figure 12, when combined with a realistic work-place subzone pattern, the assignment produces highly coherent patterns in the workplace distribution, which is achieved by prioritizing land use and using the distance only as a proxy to filter out unlikely combinations.

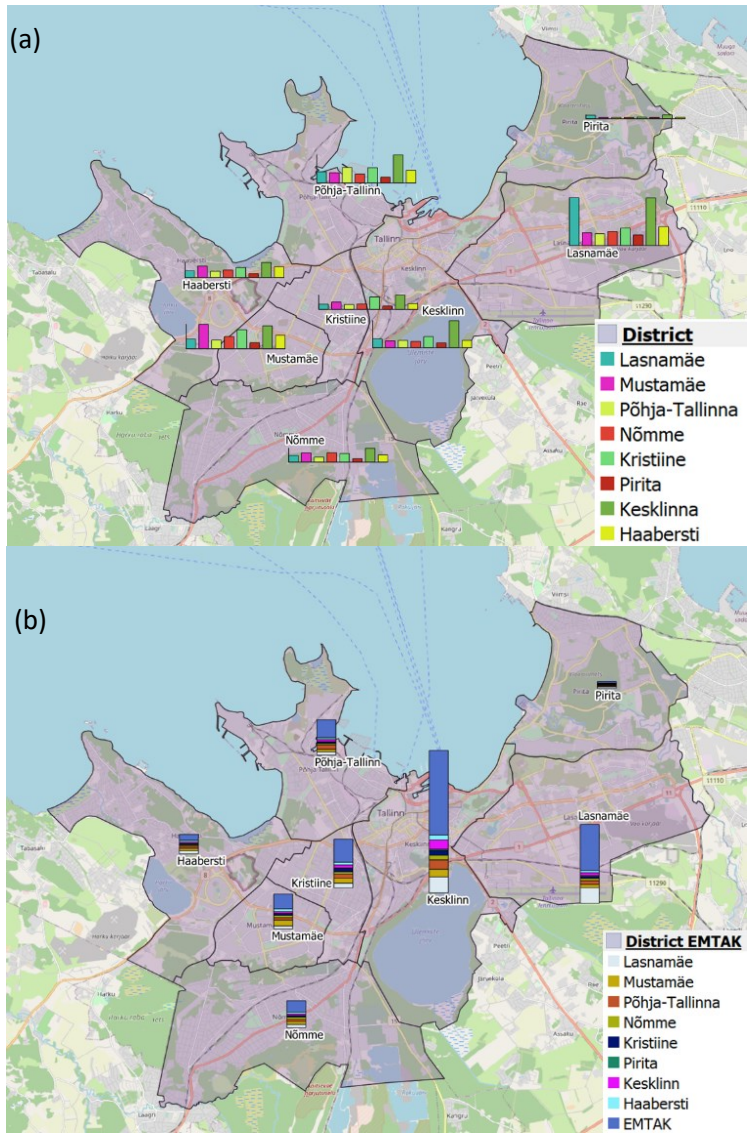

**Figure 12:** (a) Residents and their workplace district as destination; (b) Workers and their district of residence as origin

In particular, as it can be noticed from Figure 12a, the residents of Läsnamäe, for example, do work mostly in Läsnamäe and Kesklinn. The residents of Haabersti, on the other hand, tend to work in Haabersti and in Kesklinn. This captures the various trends that might be expected: the district of residence indeed attracts a fair share of workplaces while the central districts do the same. Thus, expected movements within the city are replicated while reducing skewedness to a minimum. Figure 12b aims at comparing the total of employees per district with the total per district in the resulting dataset. As it can be seen from a qualitative perspective, the assignment reproduces quite faithfully the various shares. It should be highlighted that data for a quantitative comparison are not available. The addition of the gravitational element for not coherent EMTAK fields does not skew the totals. The distribution among the cells seem to reproduce the most coherent patterns as well.

In Figure 13, the workplace cells for people residing in Nõmme and working in Põhja-Tallinna are reported. It is worth highlighting how the commuting pattern here faithfully represents the land use situation even though a gravitational model was implemented. Indeed, the darker squares in the northern part represent the zones in Põhja-Tallinna where the harbor buildings are situated, while the southern darker squares are the result of the added gravitational element that capture the areas near the city center and near the train station. Besides, this pattern is specific to commuting between Põhja-Tallinna and Nõmme, since the residence/work district distribution was kept through the EMTAK field assignment. This way, all the existing information are exploited for the assignment, with the gravitational hypothesis coming into play only when no other relevant information is available.

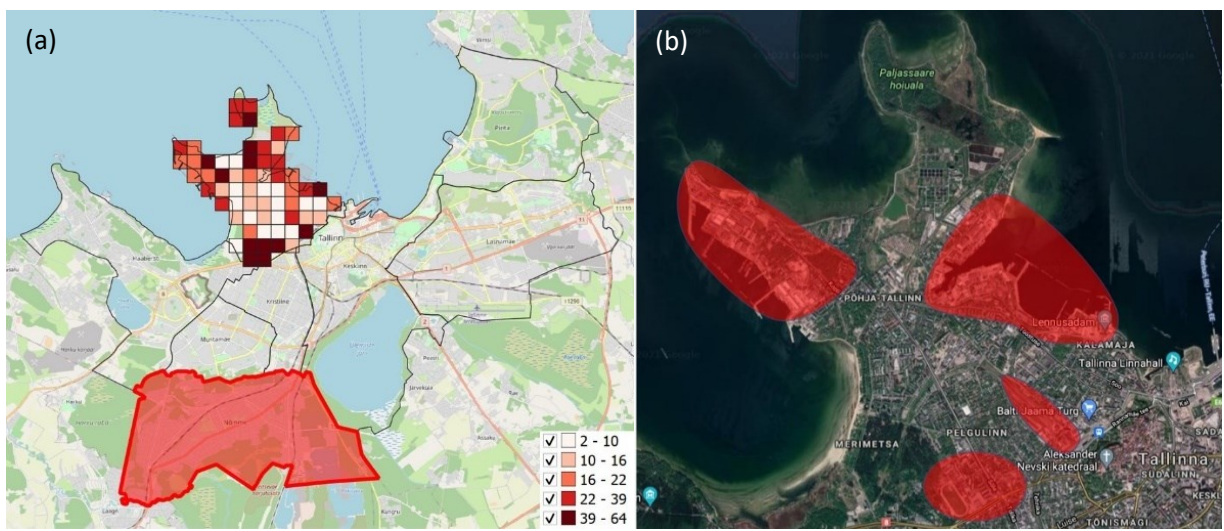

**Figure 13:** Commuting patterns – darker cells being the ones with the more workplaces; (a) from assignment; (b) from Google Maps

A final comparison is performed against (*26*), a recent study investigating commuting patterns in Tallinn by exploiting mobile phone cellular network data. The reference period of the study was May 2018 and the degree of precision was at district level. Note that the two studies are built on very different datasets and the raw data from the mobile phone network were not available to the authors. Table 1 provides the results obtained by both the proposed workplace assignment method and the ones reported in (26); one can observe that the hypotheses made about land use, EMTAK fields and distances frame the overall trends, even capturing outliers such as Pirita-Lasnamäe accounting for a 22% of the residing workforce.

**Table 1:** Residence/Workplace pattern obtained (a) from the workplace assignment and (b) from (*26*)

| (a) Synthetic Population | Mustamäe | Lasnamäe | Pohja-Tallinna | Kesklinna | Nomme | Haabersti | Kristiine | Pirita |
|---|---|---|---|---|---|---|---|---|
| Mustamäe | 0.21 | 0.08 | 0.07 | 0.20 | 0.10 | 0.11 | 0.17 | 0.05 |
| Lasnamäe | 0.07 | 0.27 | 0.06 | 0.26 | 0.08 | 0.10 | 0.10 | 0.06 |
| Pohja-Tallinna | 0.10 | 0.10 | 0.14 | 0.26 | 0.08 | 0.12 | 0.14 | 0.06 |
| Kesklinna | 0.08 | 0.11 | 0.09 | 0.34 | 0.08 | 0.09 | 0.14 | 0.06 |
| Nomme | 0.15 | 0.10 | 0.08 | 0.21 | 0.15 | 0.11 | 0.14 | 0.06 |
| Haabersti | 0.16 | 0.09 | 0.09 | 0.20 | 0.10 | 0.16 | 0.14 | 0.05 |
| Kristiine | 0.12 | 0.09 | 0.08 | 0.25 | 0.08 | 0.10 | 0.21 | 0.05 |
| Pirita | 0.07 | 0.22 | 0.08 | 0.25 | 0.08 | 0.10 | 0.11 | 0.08 |

| (b) Results from (26) | Mustamäe | Lasnamäe | Pohja-Tallinna | Kesklinna | Nomme | Haabersti | Kristiine | Pirita |
|---|---|---|---|---|---|---|---|---|
| Mustamäe | 0.36 | 0.06 | 0.08 | 0.20 | 0.04 | 0.12 | 0.12 | 0.02 |
| Lasnamäe | 0.05 | 0.40 | 0.07 | 0.32 | 0.03 | 0.04 | 0.06 | 0.03 |
| Pohja-Tallinna | 0.05 | 0.08 | 0.33 | 0.28 | 0.03 | 0.08 | 0.12 | 0.02 |
| Kesklinna | 0.06 | 0.09 | 0.10 | 0.54 | 0.03 | 0.06 | 0.09 | 0.02 |
| Nomme | 0.11 | 0.07 | 0.07 | 0.32 | 0.26 | 0.07 | 0.10 | 0.01 |
| Haabersti | 0.16 | 0.07 | 0.08 | 0.20 | 0.04 | 0.35 | 0.10 | 0.01 |
| Kristiine | 0.13 | 0.06 | 0.10 | 0.29 | 0.05 | 0.09 | 0.28 | 0.01 |
| Pirita | 0.02 | 0.22 | 0.08 | 0.31 | 0.04 | 0.04 | 0.10 | 0.20 |

**Table 2:** Difference of percentage (Delta %) between the results of this paper and (26)

| Delta % | Mustamäe | Lasnamäe | Pohja-Tallinna | Kesklinna | Nomme | Haabersti | Kristiine | Pirita |
|---|---|---|---|---|---|---|---|---|
| Mustamäe | -0.15 | 0.02 | -0.01 | 0.00 | 0.06 | -0.01 | 0.05 | 0.03 |
| Lasnamäe | 0.03 | -0.13 | -0.01 | -0.06 | 0.05 | 0.06 | 0.04 | 0.03 |
| Pohja-Tallinna | 0.05 | 0.02 | -0.19 | -0.02 | 0.05 | 0.04 | 0.02 | 0.04 |
| Kesklinna | 0.02 | 0.02 | -0.01 | -0.20 | 0.05 | 0.03 | 0.05 | 0.04 |
| Nomme | 0.04 | 0.03 | 0.01 | -0.11 | -0.11 | 0.04 | 0.04 | 0.05 |
| Haabersti | 0.00 | 0.02 | 0.01 | 0.00 | 0.06 | -0.19 | 0.04 | 0.04 |
| Kristiine | -0.01 | 0.03 | -0.02 | -0.04 | 0.03 | 0.01 | -0.07 | 0.04 |
| Pirita | 0.05 | 0.00 | 0.00 | -0.06 | 0.04 | 0.06 | 0.01 | -0.12 |

In Table 2, the difference between the percentage in Table 1 are reported. As it can be noticed, most of the differences sit below or around 5%. Still, a pattern of errors emerges for the case where workers reside and work in the same district, which are consistently higher in (26). This may have different interpretations and the two most plausible hypotheses in the authors' opinion are:

- In this study, the weight of distance should be higher for workplaces in the immediate proximity of the abitation. This would reflect a nonlinear pattern in the relevance of distance. The distance would weight more than the other factors (land use and EMTAK fields) for the cells in the immediate surroundings of the residence location.

- In (26), workplaces are identified as the most frequent cell-ID registered between 11:00 and 16:00 during working days. Cases in which these cell-IDs are the same as the residence ones are excluded. While this filtering probably captures most of homeworkers, retired people, and people with a different work schedules, the approach may fail in identifying some outliers (e.g., stay-at-home parents with a gym routine). This would result in an overestimation of the people working and residing in the same district.

The authors' hypothesis is that the truth lies probably in between, with the approach presented in this paper failing to capture some of the preferences related to a closer workplace and the approach in (*26*) possibly capturing some unintended entries due to the anonymity of the data. Finally, some small discrepancy may be due to the different reference year (2015 and 2018).

To summarise, it was possible to validate residence and workplace patterns, the spatial distribution of various household sizes, and the distribution of gender and age within the population. Thus, each step of the process was checked against the best available data and the results were deemed promising.

## 4.   Conclusions

The paper describes a systematic methodology to assign workplace anchor points to a synthetic population, by exploiting land use data and an aggregate dataset with totals of employees per NACE field, which can be employed for activity-based demand generation. The designed method is conceived as nimble, modular (i.e., not bound to any existing tool), and reliant on mostly open and/or aggregate datasets. The aim of the paper was to report the methodological steps, while highlighting the current limits both concerning the available solutions in literature and the data that may be available for most use cases. The described solution is replicable and highly transferable, with the main strength lying in its simplicity and low reliance on available data. In particular, the transferability is ensured by the fact that the proposed methodology exploits only open data in the format commonly registered by national or sovranational statistical bodies and that no foreseeable barrier may be identified a priori. For example, as mentioned in Section 3, the most important dataset exploited is the NACE one, built on the Eurostat classification with both a national equivalent in Estonia (EMTAK) and an American equivalent in the NACS classification.

While the results look encouraging and amount to a working input for ABM (they were tested on SimMobility MT), some limitations are still present and are worth being discussed. First, the lack of more disaggregated data limited the validation that could be performed. While this is indeed the issue that the paper tries to tackle, it is clear that replicating this approach to other case studies with more available data would further allow to quantify the achieved degree of precision. Another limitation of the paper is that it exploits the weights reported in (*17*), conceived for a prototype city. Future development could investigate the presented method applied to a case study in which detailed land use data are recorded to perform a calibration of said weights. Besides, other open mobility data, such as public transport lines and skim matrixes, could improve precision, i.e., by skewing the weights of certain cells close to mobility hubs.

Future development may also concern the integration of the public transport stops in the weighting land use factors or focus on the weighting of the distance factor in the immediate vicinities of the residence, to investigate the discrepancies highlighted in Section 3.

The resulting dataset is detailed to be exploitable by fellow researchers for Activity-Based modeling (or any other research direction) since it is shared as open source.

## Acknowledgements

This research is funded by the FINEST Twins Center of Excellence (H2020 European Union funding for Research & Innovation grant 856602).

The authors would like to thank Prof. Dago Antov from Taltech for sharing the travel survey exploited in this work. Moreover, the authors are grateful to the Tallinn Municipality and to all the related public bodies who supported this research by sharing relevant data.

## COMPETING INTERESTS STATEMENT

The authors declare no competing interests.

## Supplementary material

The Tallinn syntentic population dataset is available at:

[https://github.com/Angelo3452/Tallinn-Synthetic-Population](https://github.com/Angelo3452/Tallinn-Synthetic-Population)

It is provided as open source and licensed under Creative Commons — Attribution 4.0 International — CC BY 4.0. Its structure and the provided variables are there described, while additional relevant distributions are reported. Two examples implementing the methodology described in this paper are also included, written in the programming language R.

**Appendix**

| Data | Type of data | Source | Usage | Public/Private |
|---|---|---|---|---|
| Household structure | Survey data | Survey from Taltech | Synthetic Population | Private |
| Age x Gender distribution | Statistical margin | Statistical Yearbook of Tallinn | Synthetic Population | Public |
| Household size x District distribution | Statistical margin | Statistical Yearbook of Tallinn | Synthetic Population | Public |
| Population x Subdistrict | Statistical margin | Statistical Yearbook of Tallinn | Synthetic Population | Public |
| Car Ownership x Household size | Probability distribution | Survey from Taltech | Synthetic Population | Private |
| Income per family member x Subdistrict | Distribution | Municipality of Tallinn | Validation | Upon request |
| Residential buildings x cell [m$^2$] | Land Use | Tallinn Geoportal | Weight assignment | Public |
| Manufacturing and industrial buildings x cell [m$^2$] | Land Use | Tallinn Geoportal | Weight assignment | Public |
| Service and office buildings x cell [m$^2$] | Land Use | Tallinn Geoportal | Weight assignment | Public |

| Enrollment x educational building | Assignment | Ehis database[7] | Spatial assignment | Public |
|---|---|---|---|---|
| Location of each educational building | Assignment | Ehis database | Spatial assignment | Public |
| Classification of each educational building | Assignment | Ehis database | Spatial assignment | Public |
| District of residence x enrollments in each district | Assignment | Ehis database | Spatial assignment | Upon request |
| Number of employees x District x EMTAK field | Assignment | RIK | Spatial assignment | Publicly available for a fee |
| Gender, Age and District of residence x Occupation | Assignment | Census | Spatial assignment | Public |
| Occupation x Emtak field | Assignment | Census | Spatial assignment | Public |
| Household structure | Survey data | Survey from Taltech | Synthetic Population | Private |
| Age x Gender distribution | Statistical margin | Statistical Yearbook of Tallinn | Synthetic Population | Public |
| Household size x District distribution | Statistical margin | Statistical Yearbook of Tallinn | Synthetic Population | Public |

---

[7] https://enda.ehis.ee/avalik/avalik/oppeasutus/OppeasutusOtsi.faces : Database of educational institutions and enrollment statistics